\documentclass{aa}
\usepackage{graphicx}
\input{psfig}
\begin{document}

\title{Atomic diffusion in star models of type earlier than G}
\author{P. Morel, F. Th\'evenin}
\institute{
D\'epartement Cassini, UMR CNRS 6529, Observatoire de la C\^ote 
d'Azur, BP 4229, 06304 Nice CEDEX 4, France.}

\offprints{P. Morel}
\mail{Pierre.Morel@obs-nice.fr}

\date{Received 7 February 2002 / Accepted 15 May 2002}

\abstract{ 
We introduce the mixing
resulting from the radiative diffusivity associated with the radiative viscosity
in the calculation of stellar evolution models.
We find that the radiative diffusivity significantly diminishes
the efficiency of the gravitational settling in the external layers of
stellar models corresponding to types earlier than $\approx $\,G.
The surface abundances of chemical species predicted by the models are
successfully compared with the abundances 
determined in members of the Hyades open cluster.
Our modeling depends on an efficiency parameter, which is evaluated to a
value close to unity, that we calibrate in this study.
}

\maketitle

\section{Introduction}\label{sec:int}
Microscopic diffusion, sometimes named ``atomic'' or ``element''
 diffusion, when used
in the computation of models of main sequence stars produces
(see Chaboyer et al.~\cite{cfn01}
for an abridged revue) a change in the surface abundances from
their primordial values and
a shift of the evolutionary tracks towards lower luminosities
and temperatures (Morel \& Baglin~\cite{mb99}),
consequently increasing the expected age of binaries and clusters.
For the Sun, the insertion of microscopic diffusion has
spectacularly improved the agreement between the theoretical
and the ``seismic'' model (e.g. Christensen-Dalsgaard et al.~\cite{cda96}).
Indeed, for low mass stars and for population~II stars, atomic diffusion
eliminates, at least partially, discrepancies between observations
and theoretical models (Morel \& Baglin~\cite{mb99};
Lebreton et al.~\cite{lpcbf99}; Cayrel et al.~\cite{clm99};
Chaboyer et al.~\cite{cfn01}; Salaris \& Weiss~\cite{sw01}).

For stellar models with mass larger than $\approx 1.4\,M_\odot$,
the use of microscopic diffusion {\em alone} produces
at the surface an important depletion of helium and
heavy elements and a concomitant enhancement of the hydrogen content.
As an example, for a $2\,M_{\odot}$ model computed with atomic diffusion, 
at the age of 20\,Myr
the surface mass fraction of hydrogen increased from $X=0.70$ to $X\goa0.98$,
while helium and heavy elements are strongly depleted.
This strong depletion, which
increases with the mass of the star, is not
observed either by Varenne \& Monier~(\cite{vm99})
in stars belonging to the Hyades open cluster or in
OB associations (Daflon et al.~\cite{dcbs00}) and in the Orion association
(Cunha \& Lambert~\cite{cl94}).

There is no possibility of
observing helium in main sequence stars of intermediate mass to check its
depletion; such stars do not show low surface metallicities.
This large helium depletion is not likely to be
observed in main sequence B-stars,
as supported by their spectral classification which is mainly
based on the relative strength of the helium spectral lines.

The large efficiency of the atomic diffusion in the outer layers results
from the large temperature and pressure gradients.
For models of stars with types later than $\approx$\,G, i.e.
$M_\star\loa1.2\,M_\odot$, the external convection zone acts as a
``reservoir'' which inhibits atomic diffusion. On the contrary, for stars
with masses larger than $\goa1.4\,M_\odot$,
the external helium and hydrogen convection zones are so tiny that the
reservoir is not large enough to be efficient.

Such misleading physics in the calculation of microscopic
diffusion coefficients (e.g. Cox et al.~\cite{cgk89};
Proffitt \& Michaud~\cite{pm91}; Michaud \& Proffit~\cite{mp93}; Thoul et
al.~\cite{tbl94}) are very unlikely owing to the high degree of sophistication 
of the method based on the kinematical theory of gases
(Burgers~\cite{b69}) and also as seen by its success when applied to the Sun.
Clearly, a physical process which inhibits the efficiency of microscopic
diffusion is lacking.

Among physical processes acting against gravity, recent developments were
focused on:

\begin{itemize}

\item The radiative accelerations, which are efficient only in the case of
chemical species with a large charge, e.g. iron. As such ``metals'' have 
small relative abundances, the radiative accelerations
are not efficient limiting the
sedimentation of helium and heavy elements as a whole
(e.g. Alecian et al.~\cite{amt89}; Turcotte et al.~\cite{trmir98}).

\item The turbulent mixing generated by the shear
created by the differential rotation between the solar convection zone and the
radiative core (Richard et al.~\cite{rvcd96};
Gabriel~\cite{g97}; Brun et al.~\cite{btz99}).
It has been recently claimed (Schatzman et al.~\cite{szm00})
that the shear instability invoked to produce this turbulent mixing
only concerns a narrow region and may be not as efficient as expected.

\item The rotation-induced mixing accounting for the transport
of angular momentum explains well the hot side of the Li dip in the Hyades
(Talon \& Charbonnel~\cite{tc98}).

\item The use of a hypothetical mixed reservoir
with an ad-hoc adjusted size (e.g. Turcotte et al.~\cite{trm98}). 

\item The refinement of the physical description. In
the calculations of G to A type star models
(Turcotte et al.~\cite{trmir98},~\cite{trm98};
Richer et al.~\cite{rmt00}), the local mean Rosseland opacity is updated
at each time step, according to local
changes in the density and temperature, but also
in the genuine mass fractions of chemical species. In some favorable
situations the authors have obtained the formation of an
``iron'' mixed convective zone 
that  inhibits the efficiency of the atomic
diffusion. Even with this noticeable improvement of the modeling,
several difficulties remain;
in particular the models do not reproduce the
abundances of star members of the Hyades open cluster as observed by
Varenne \& Monier~(\cite{vm99}); their figures 5 to 8 for elements
from C to Ba only reveal a mild depletion by a factor
$\approx3$ for $6\,600$\,K$\leq T_{\rm eff}\leq 7\,800$\,K. Also
the theoretical strong depletions predicted by Turcotte et al.~(\cite{trm98})
for effective temperatures larger than $6\,600$\,K are not observed.

\item The influence of the stellar mass-loss described 
phenomenologically (Chaboyer et al.~\cite{cdg99}) 
in the modeling of Procyon\,A.

\end{itemize}

\medskip
In this exploratory work, we focus on the role of
 the radiative diffusivity generated by
the photon-ion collisions that is not presently taken into account in
the microscopic diffusion coefficients.
As we shall see, the photon-ion collisions are an efficient physical
process that inhibits the large sedimentation of helium and heavy elements
in outer layers of main sequence star models with types
earlier than $\approx$\,G.

\bigskip
In Sect.~\ref{sec:trad} we present a simplified phenomenological
model of the radiative
diffusivity resulting from photon-ion collisions. Section~\ref{sec:mod} briefly
describes the theoretical 
framework used for constructing stellar evolution models.
In Sect.~\ref{sec:estim} we estimate the radiative
diffusivity parameter connecting the
radiative kinetic viscosity to the radiative diffusivity; our investigation is
based mainly on observed abundances of C, O, Fe and Mg in stars
belonging to the Hyades open cluster as reported by
Varenne \& Monier~(\cite{vm99}).
Illustrations are given in Sect.~\ref{sec:ill} and,
in Sect.~\ref{sec:con}, we summarize and discuss
the main results of the present investigation.

\begin{figure}
 \centerline{
 \psfig{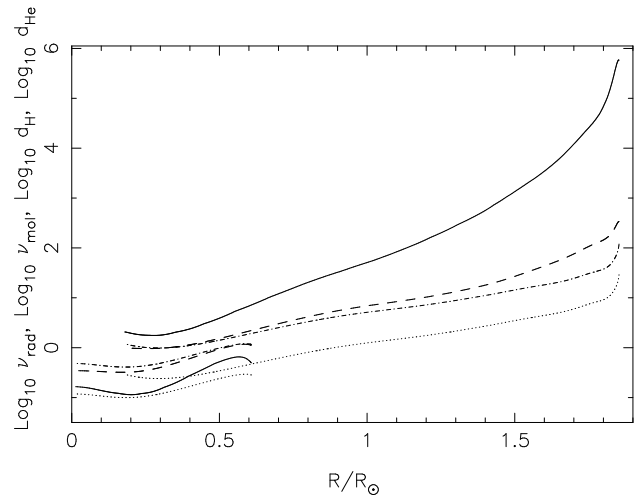}} 
 \caption[]{Profiles of kinetic radiative viscosity
 $\nu_{\rm rad}$ (heavy full),
 kinetic molecular viscosity $\nu_{\rm mol}$ (heavy dashed),
hydrogen diffusivity $d_{\rm H}$ (dot-dash-dot-dash)
 and helium diffusivity $d_{\rm He}$ (dotted)
with respect to radius in the radiative zones
of a $1\,M_\odot$ ($R\leq 0.61\,R_\odot$) and in a 
$1.8\,M_\odot$ stellar models at age 700\,Myr. The kinematic molecular and
radiative viscosities 
and the diffusivities are in c.g.s. (cm$^2$\,s$^{-1}$) units.
 }\label{fig:numol_rad}
\end{figure}

\section{Radiative diffusivity}\label{sec:trad}
Apparently, the radiative diffusivity, first used by
Baglin~(\cite{b72}), has not been introduced for intermediate stellar mass
models, although implemented in the evolution of models of
larger mass (Heger et al.~\cite{hlw00}).
According to the standard description of the micro-physics of gases
(e.g. Burgers~\cite{b69}; Mihalas \& Weibel-Mihalas~\cite{mm84}, Chap.~1)
the viscous stresses due to electric forces tie up
the constituent particles, electrons and ions.
In an astrophysical plasma with radiation, the kinematic viscosity $\nu$
has two components:
\[\nu=\nu_{\rm mol}+\nu_{\rm rad}.\]
The kinematic molecular viscosity, $\nu_{\rm mol}$, results from
energy exchanges between thermal collisions leading to excitation and
ionization of atoms and ions. Thus, the thermal collisions generate
a ``molecular mixing'' resulting from the molecular diffusivity associated
with $\nu_{\rm mol}$.
The kinematic molecular viscosity is approximated by (Schatzman~\cite{s77}): 
\begin{equation}\label{eq:nu_mol}
\nu_{\rm mol}\simeq2.21\,10^{-15}\frac{T^{\frac52}}\rho
\frac{1+7X}{3\ln\Lambda},
\end{equation}
for a mixture of hydrogen and helium, assuming that the heavy elements
have a negligible effect.
$T$ is the temperature, $\rho$ is the density, $X$ is the hydrogen mass fraction
and $\ln\Lambda$ is the so-called Coulomb logarithm (e.g. Michaud \&
Proffit~\cite{mp93}).
With the more elaborate description
based on the kinetic theory of gases (Burgers~\cite{b69}), 
the molecular diffusion has two components: 
the isotropic one, being the molecular diffusivity that generates a
mixing and the anisotropic one, responsible for the gravitational
sedimentation in stars.
 
The kinematic radiative viscosity $\nu_{\rm rad}$ arises because photons
deposit their momentum in the fluid element into which they are absorbed. Thus,
the radiative collisions generate a ``radiative mixing'' resulting from
the radiative diffusivity associated with $\nu_{\rm rad}$. 
In the optically thick medium 
the kinematic radiative viscosity is expressed as
(Thomas~\cite{t30}; Mihalas \& Weibel-Mihalas~\cite{mm84}, p. 461-472):
\begin{equation}\label{eq:nu_rad}
\nu_{\rm rad}\equiv \frac4{15}\frac{aT^4}{c\kappa\rho^2},
\end{equation}
where $a$ is the radiation density constant, $c$ is the
speed of light in vacuum and $\kappa$ is the mean Rosseland opacity.

As already emphasized by several authors more than seventy years ago
(e.g. Milne~\cite{m30}), the radiative viscosity is large
in the external layers of stars (Baglin~\cite{b72}).
Figure~\ref{fig:numol_rad} shows the profiles, with respect
to radius, of $\nu_{\rm mol}$ and $\nu_{\rm rad}$
in the radiative zones of two main sequence stellar models of
respectively $1.0\,M_\odot$ and $1.8\,M_\odot$ at age 700\,My.
In the radiative zone of the $1\,M_\odot$ model ($R\leq 0.61\,R_\odot$),
the total kinematic
viscosity mainly results from the kinematic molecular viscosity while in the 
radiative outermost layers of the $1.8\,M_\odot$ model, the kinematic radiative
viscosity overcomes the kinematic molecular viscosity by several
orders of magnitude. The large value of the radiative viscosity in
outer layers of stars is associated with a large 
radiative diffusivity that
makes a significant contribution to the mixing.
Like the diffusion due to molecular
collisions, the radiative diffusion has an isotropic component,
the radiative diffusivity generating
mixing, and an anisotropic component causing the radiative
accelerations that act against gravity (Alecian~\cite{a96}).

For stars with mass $M_\star\la1.2M_\odot$ the effect of radiative diffusion
is swallowed up by the mixing in the external convection zone, as in the
Sun. For stars of types earlier than G, with increasing mass (i.e. effective
temperature) the external convective zones become smaller
and, in outer layers, the radiative diffusivity
is a process which acts more and more efficiently against the gravitational
sedimentation.

In the most elaborated stellar evolution codes,
the coefficients of microscopic diffusion are derived using the
Burgers~(\cite{b69}) formalism (e.g. Cox et al.~\cite{cgk89};
Proffitt \& Michaud~\cite{pm91}; Michaud \& Proffit~\cite{mp93}; Thoul et
al.~\cite{tbl94}). But, 
in the standard framework of the calculation of collision integrals
and resistance coefficients
(Burgers~\cite{b69}; Paquette et al.~\cite{ppfm86}),
only the ion-ion and electron-ion collisions
are taken into account and the derived diffusivity
corresponds only to $\nu_{\rm mol}$. 

Figure~\ref{fig:numol_rad} also shows
the profiles of the hydrogen $d_{\rm H}$ and helium $d_{\rm He}$
molecular diffusivities. Note that 
$d_{\rm H}$, $d_{\rm He}$ and $\nu_{\rm mol}$ have similar profiles.
This behavior
suggests a linear relationship, with a dimensionless coefficient close to unity
between the molecular diffusivity and the kinetic molecular viscosity.

\medskip
It is out of the scope of this exploratory work to compute the
diffusion coefficients including both the thermal and
the radiative collisions according to
the Burgers' (\cite{b69}, Sect.~18 \& 60) formalism.
We express the
diffusivity associated whith $\nu_{\rm rad}$ by a diffusion 
coefficient proportional to the kinematic radiative viscosity:
\[d_{\rm rad}\equiv D_{\rm R}\,\nu_{\rm rad}. \]
The efficiency factor, $D_{\rm R}$, is a dimensionless
phenomenological parameter
to be calibrated. Hereafter, we shall refer to $D_{\rm R}$ as the
``radiative diffusivity parameter''.
We assume that $D_{\rm R}$ is constant
even if, {\em a priori}, there is no reason for a
unique value to apply for all atoms and ions of all chemical species
and in all stellar conditions.

\begin{figure*}
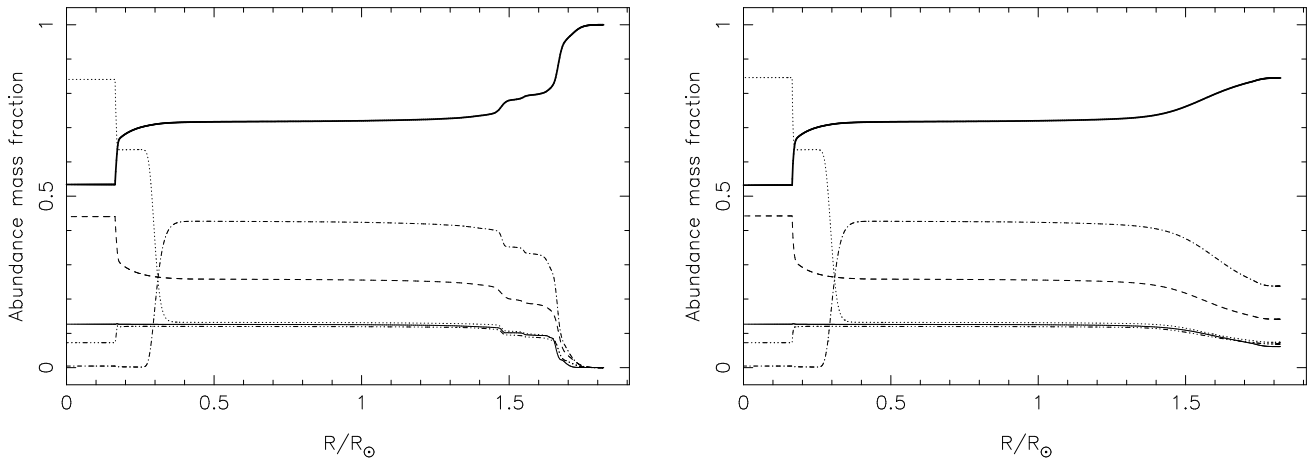

\centerline{
\hbox{\psfig{figure=2356f2.eps,height=6.cm,angle=270}
\hspace{0.5truecm}
\psfig{figure=2356f3.eps,height=6.cm,angle=270}}
}
\caption{Profiles, with respect to radius, of abundance mass fractions in a
$1.8\,M_\odot$ ``Hyades'' stellar model at age $570$\,Myr computed with
$D_{\rm R}=0$ (left) and $D_{\rm R}=1$ (right). $\rm H\times1$ (heavy full),
$\rm He\times1$ (dashed), $\rm C\times100$ (dot-dash-dot-dash),
$\rm N\times100$ (dotted), $\rm O\times100$ (dash-dot-dot-dot-dash),
$\rm Fe\times100$ (thin full).
}\label{fig:rm}
\end{figure*}

To illustrate the effect of the radiative diffusivity, using the
results of Sect.~\ref{sec:ill}, Fig.~\ref{fig:rm} shows the abundance profiles
with respect to radius in a
$1.8\,M_\odot$ main sequence stellar model at age $570$\,Myr computed
respectively with ($D_{\rm R}=1.0$)
and without ($D_{\rm R}=0.0$) radiative diffusivity. Note the small surface
abundances
of metals predicted by the model computed without radiative diffusivity
and the realistic values obtained with $D_{\rm R}=1.0$.

The simplified description of the radiative diffusivity introduced in this
paper is phenomenological and preliminary.
Particularly it ignores the ionization stage of elements
which could modify the efficiency of the radiative diffusivity
of chemical species under consideration.
It obviously needs further extended development based
on the kinetic theory of gases with radiation.

\section{Computation of models and physical inputs}\label{sec:mod}
\paragraph{Stellar modeling:}
Stellar models have been computed using the {\sc cesam} code (Morel~\cite{m97}).
The physics employed is the same as in
Morel et al.~(\cite{mpltb00}).
The stellar evolution sequences are initialized with homogeneous 
pre-main sequence models.
We follow in detail the abundances of all chemical
species of our diffusion network.
The set of evolutionary differential equations fulfilled by
the abundances of the isotopes is numerically integrated
using a finite element method\footnote{The derivation of the equations and the
algorithms are detailed in the {\it ``Notice de
CESAM''} and available on request on the WEB site
{\tt http://www.obs-nice.fr/morel/CESAM.html}.}. The microscopic diffusion
coefficients
are derived from the set of Burgers'~(\cite{b69}, formula 18.1-18.2)
flow equations. The charge of a given isotope is taken as the
average\footnote{Weighted by the ionization rates.}
charge over all its ionization states.
The collision integrals and resistance coefficients 
are taken from the tabulations of Paquette et al.~(\cite{ppfm86}). For each
ion, the coefficient of radiative diffusivity
is added to the coefficient of molecular diffusivity.

\paragraph{Input physics:}
The {\sc eff} equation of state (Eggleton et al. \cite{eff73}) was used,
together with opacities determined by Rogers et al.~(\cite{rsi96})
using the solar mixture of Grevesse \& Noels~(\cite{gn93}),
complemented at low temperatures by
Alexander \& Ferguson~(\cite{af94}) opacities.
In the convection zones the mixing is insured by a large turbulent diffusion
coefficient and the temperature gradient is
computed according to the B\"{o}hm-Vitense~(\cite{b58}) convection theory;
the mixing-length parameter is fixed to the value in the standard solar
model computed with the same physics\footnote{But the inefficient
radiative diffusivity, see Sect.~\ref{sec:sol}.}.
The atmosphere is restored using Hopf's atmospheric law (Mihalas~\cite{m78}).
The element diffusion is not computed in the atmosphere where
the chemical composition is fixed at the value in the outermost shell
of the envelope.   
 
\paragraph{Nuclear network:} The relevant nuclear reaction rates
are taken from the {\sc nacre} compilation (Angulo et al.~\cite{aar99}).
The nuclear network contains the isotopes:
\element[][1]{H},
\element[][2]{H},
\element[][3]{He},
\element[][4]{He},
\element[][7]{Li},
\element[][7]{Be},
\element[][12]{C},
\element[][13]{C},
\element[][14]{N},
\element[][15]{N},
\element[][16]{O} and
\element[][17]{O}.
They are processed by the {\sc pp+cno}
thermonuclear reactions relevant for the main sequence evolution.
The elements
\element[][2]{H} and \element[][7]{Be} are set at
equilibrium.

\paragraph{Diffusion network:}
Along the main sequence the chemical species,
heavier than \element[][17]{O}, are not nuclearly processed. However,
their abundances vary with respect to time owing to diffusion processes, 
contributing to changes of $Z$, thus generating opacity variations.
Morel et al.~(\cite{mpb97}) introduced a dummy ``mean metal''
to approximate the mean opacity changes caused by the variations of $Z$
resulting from microscopic diffusion.
To mimic more closely the effects of low and high charged chemical species,
we use here two dummy mean metals. 
The first simulates the metals heavier than potassium.
For the solar mixture of Grevesse \& Noels~(\cite{gn93}) used so far, their
mean\footnote{Weighted by number.} charge
and mean atomic weight are respectively 26 and 55.
Therefore, a dummy ``iron'', labeled hereafter \{\element[][55]{Fe}\},
is well representative of
the heaviest chemical species.
Indeed, the second dummy metal simulates the isotopes with atomic
 numbers from 9 to 20,
and also complements the mixture. Its charge and atomic mass are
respectively 12 and 24, corresponding to a ``dummy'' magnesium
\{\element[][24]{Mg}\}. For the sake of simplicity we shall approximate
\{\element[][55]{Fe}\} and \{\element[][24]{Mg}\} with respectively
the iron and the magnesium abundances measured spectroscopically.

\paragraph{Initial abundances:}
To fulfill
the basic relationship $X_{\rm ini}+Y_{\rm ini}+Z_{\rm ini}\equiv1$,
the initial abundances of isotopes are derived  (1) from
the isotopic ratios of nuclides (Anders \& Grevesse~\cite{ag89}), (2) from
the initial mass fraction of helium,
$Y_{\rm ini}\equiv \element[][3]{He_{ini}}+\element[][4]{He_{ini}}$,
and (3) from
the initial mass fraction of heavy elements relative to hydrogen
$(\frac ZX)_{\rm ini}$. As deuteron
 is assumed at equilibrium the initial amount of
\element[][3]{He} is enhanced by the initial meteoritic
amount of deuteron (Geiss \& Reeves~\cite{gr72}).

Failing anything better, we use the meteoritic values
of Anders \& Grevesse~(\cite{ag89}) for
the initial fractions, by number, of metals within $Z$:
$\rm ^{12}C+^{13}C=0.2455$,
$\rm ^{14}N+^{15}N=0.06458$,
$\rm ^{16}O+^{17}O=0.5130$.
From these data one infers the initial abundances for the dummy metals
$\rm\{^{24}Mg\}=0.1631$,
$\rm\{^{55}Fe\}=0.01563$. 
For the initial isotopic ratio (by number) we use
\element[][3]{He}/\element[][4]{He}$=1.1\,10^{-4}$
(Gautier \& Morel~\cite{gm97}).

\bigskip
For any chemical species $\cal X$ we shall refer to
$[\frac{\cal X}{\rm H}]$ as the quantity:
\begin{equation}\label{eq:xsh}
\rm{[\frac{\cal X}{\rm H}]} \equiv \log\left(\frac{\cal X}{\rm H}\right)_t
- \log\left(\frac{\cal X}{\rm H}\right)_{t=0},
\end{equation}
which is a measurement of the depletion of $\cal X$ at the age $t$
under consideration.

\section{Constrains on the radiative diffusivity parameter.}\label{sec:estim}
The radiative diffusivity parameter $D_{\rm R}$ is estimated by
comparing relevant
observations to theoretical results.
Firstly, owing to the high accuracy of helioseismological observations,
some insights can be gained using the standard solar model
despite the inhibition, by the convective mixing, of
the microscopic diffusion in the outer layers. Secondly,
the most sensitive observations, relevant to the atomic diffusion, are
the surface abundances of chemical species of members
of well-observed open clusters.
When the age of the cluster is not older than $\approx1$\,Gyr,
the effects of element diffusion remain small for
 $\approx$\,G dwarfs and the
initial abundances can be reasonably inferred. For intermediate mass stars
we shall estimate the value of the radiative diffusivity parameter using the
observational material of the Hyades open cluster
and, for early type stars, we shall use observations of Orion association.

\paragraph{Solar models:}\label{sec:sol}
In the external layers of a solar model, the radiative diffusivity
is swallowed up by the
convective mixing. Beneath the convection zone, where precise informations on
the internal structure result from helioseismology, the effects of
radiative diffusivity should be
compatible with the relevant properties of the
nowadays very precise standard solar model. This assertion 
leads to an estimate of an upper limit for $D_{\rm R}$.

\begin{figure}
\centerline{
\psfig{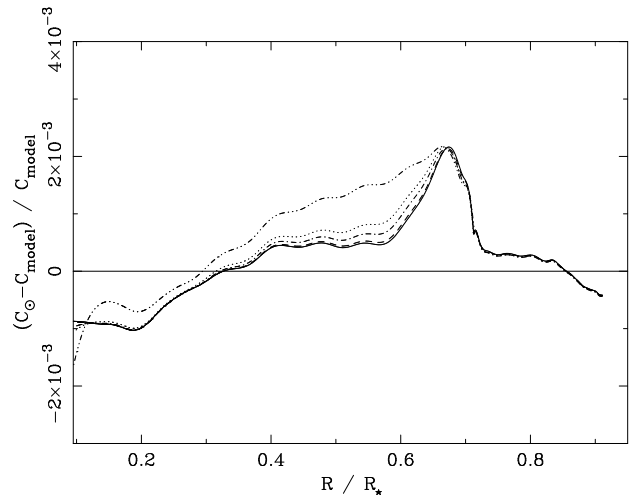}}
\caption[]{Relative differences in sound velocities between the Sun and
calibrated solar models with $D_{\rm R}=0,\, 1,\,5,\,10,\,50$  respectively:
full, dashed, dot-dash-dot-dash, dotted, dash-dot-dot-dot-dash. 
}\label{fig:vson}
\end{figure}

We have computed calibrated solar models
for the following values of the radiative diffusion parameter
$D_{\rm R}=0,\,1,\,5,\,10,\,50$.
To compare solar models with helioseismological observations, we
use improved physical data and standard solar initial chemical composition
(details can be found in Morel et al.~\cite{mpb97}).
In the range $0.1\,R_\odot\la R\la 0.9\,R_\odot$ where the inversions
of the helioseismic data are reliable,
Fig.~\ref{fig:vson} shows the relative differences between
the sound speed in the calibrated solar models
and the seismic sound speed ``experimental'' results of Turck-Chi\`eze et
al.~(\cite{tbb97}).
For the models computed with
$D_{\rm R}\loa10$, the relative discrepancy between the sound speed in the
Sun and in the model are not larger than a few $10^{-3}$. 
Table~\ref{tab:sol} shows that the theoretical values
of standard solar characteristics are all compatible with observations;
nonetheless, these data are not strongly dependant on $D_{\rm R}$.

From Fig.~\ref{fig:vson}, we estimate that a 
reasonable upper limit for the radiative diffusivity parameter
is close to $\lceil D_{\rm R}\rceil\loa10$.
\begin{table}
\caption[]{Depth of the convection zone (solar unit),
surface isotopic ratio of \element[][3]{He} versus \element[][4]{He},
surface helium mass fraction
$\rm Y_s$, small difference $\delta\nu_{02}$
($\mu$\,Hz) and $\rm [Li]_s$ lithium abundance in dex ($\rm H\equiv12$)
in calibrated solar models computed with different values for the
radiative diffusivity parameter $D_{\rm R}$. 
}\label{tab:sol}
\begin{tabular}{llllllllllll} \hline \\
$D_{\rm R}$&$R_{\rm zc}$ &$\rm [\frac{^3He}{^4He}]_s$ &$\rm Y_s$&$\delta\nu_{02}$&$\rm [Li]_s$\\
\\  \hline \\
$0$  &$0.7127$ &$4.343$ &$0.2430$ &$9.09$ &$2.603$\\
$1$  &$0.7128$ &$4.345$ &$0.2430$ &$9.10$ &$2.609$\\
$5$  &$0.7129$ &$4.353$ &$0.2431$ &$9.12$ &$2.608$\\
$10$ &$0.7131$ &$4.359$ &$0.2433$ &$9.15$ &$2.607$\\
$50$ &$0.7142$ &$4.381$ &$0.2448$ &$9.32$ &$2.502$\\
\\  \hline \\
\multicolumn{3}{l}{Observations} & \multicolumn{3}{l}{References} \\
\multicolumn{3}{l}{$R_{\rm zc}=0.713\pm0.001$} &\multicolumn{3}{l}{Basu \& Antia~(\cite{ba95})} \\
\multicolumn{3}{l}{${\rm [\frac{^3He}{^4He}]_s}=4.4\pm 0.4$} & \multicolumn{3}{l}{Bodmer et al.~(\cite{bbg95})} \\
\multicolumn{3}{l}{${\rm Y_s}=0.240\pm0.009$} & \multicolumn{3}{l}{Basu~(\cite{b97})} \\
\multicolumn{3}{l}{$\delta\nu_{02}=9.01\pm0.10$} & \multicolumn{3}{l}{Grec et al.~(\cite{gtl97})} \\
\end{tabular}
\end{table}

\paragraph{Massive stars:}
Massive stars, belonging to OB associations,
provide a test of efficiency for the radiative diffusivity.
For stars having masses from 7 to 17\,$M_\odot$ observations do not reveal
any chemical abundances depletion with respect
to the effective temperature, i.e. masses, neither for
C, N and O nor for heavier elements, as it is 
shown in Fig.~3 of Daflon et al.~(\cite{dcbs00}) or in Fig.~17 of
Cunha \& Lambert~(\cite{cl94}). These recent star-forming regions
are younger than 20\,Myr and most of the stars analysed in
Cunha \& Lambert~(\cite{cl94}) have an age supposed to be around 7\,Myr.

In models of such massive stars including element diffusion {\em alone}
($D_{\rm R}=0.$), helium and metals empty the outer layers 
in an evolutionary time shorter than $\approx10$\,Myr. In these models
the outer hydrogen and helium convection zones are so
tiny that the convective mixing is ineffective. Resulting from the 
large pressure and temperature gradients,
the element diffusion is very efficient, leading to
a large sedimentation of helium and heavy elements.
The inclusion of the radiative diffusivity generates a mixing which
avoids this unrealistic result.
Due to the large effective temperature ($T_{\rm eff}\goa10\,000$\,K),
the kinematic radiative viscosity is large in the outer layers,
see Eq.~(\ref{eq:nu_rad}), and so
the radiative diffusivity; that generates an efficient mixing
inhibiting the large efficiency of the
element diffusion and allows one to reproduce the observed absence of
depletion of helium and metals.

To estimate $D_{\rm M}$, adopting a mean age of 7\,Myr,
we have computed the evolved abundances of C, N, O and Fe
for stellar masses from 7 to $17\,M_\odot$ for different values of the
radiative diffusivity parameter.
In these early type star models the radiative diffusivity inhibits the
efficiency of the element diffusion as soon as $D_{\rm R}$ is larger
than the lower limit $\lfloor D_{\rm R}\rfloor=0.05$.

\paragraph{Hyades:}\label{sec:hay}
The Hyades cluster is a nice laboratory for testing stellar structure and
evolution theory because of its numerous well analysed stars and its
well measured distance. We have computed star models 
taking into account the atomic diffusion including radiative diffusivity
for different values of $D_{\rm R}$.
We refer to Lebreton et al.~(\cite{lfl01}) for the cluster parameters.
We adopt a metallicity of ${\rm [\frac{Fe}H]}=+0.14 \pm 0.03$\,dex
and an age of $550\pm50$\,Myr as our models are computed without
overshoot of the convective core. Alike for the Sun,
a typical G dwarf Hyades member of $1.1\,M_\odot$ is not affected
by the radiative diffusivity but, owing to the element diffusion,
the initial helium mass fraction needs to be enhanced up
to $Y_{\rm ini}=0.259$ to ends at $Y=0.255$ after $550$\,Myr
of evolution. Also, the initial metallicity needs
to be $\rm [Fe/H]_{ini}=0.147$\,dex to reach
 $\rm [Fe/H]=0.140$\,dex at the age of Hyades.
For our ``Hyades'' models we use the following initial mass fractions
(in dex with respect to $\rm H\equiv12$):
\element[][3]{He}=8.056,
\element[][4]{He}=11.56,
\element[][12]{C}=9.738,
\element[][13]{C}=7.818,
\element[][14]{N}=9.299,
\element[][15]{N}=6.896,
\element[][16]{O}=10.24,
\element[][17]{O}=6.847,
\{\element[][24]{Mg}\}=9.416,
\{\element[][55]{Fe}\}=9.838.

\begin{figure*}
 \centerline{
 \psfig{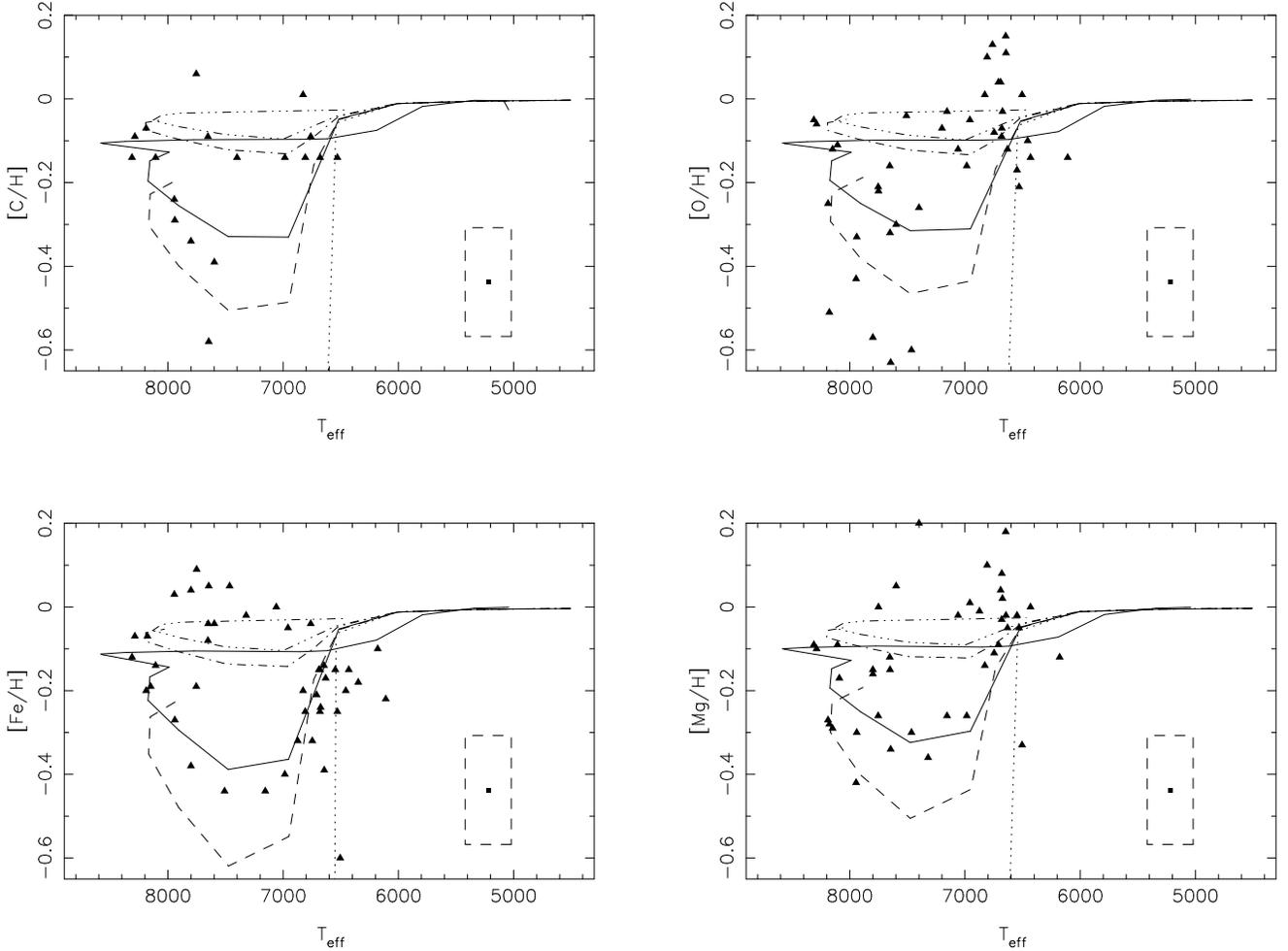}}
 \caption[]{Surface abundances (in dex) with respect to effective temperature
 of C, O, Fe and Mg for a sample of
 ``Hyades'' stellar models at the age of $570$\,My computed with
 $D_{\rm R}=0.0$ (dotted),
 $D_{\rm R}=0.5$ (dashed), $D_{\rm R}=1.0$ (full),
 $D_{\rm R}=5.0$ (dash-dot-dash-dot) and 
 $D_{\rm R}=9.0$ (dash-dot-dot-dot-dash). The triangles correspond to
 the observations of Varenne \& Monier~(\cite{vm99}) corrected
 from the metallicity of Hyades. Dashed rectangles delimit the uncertainty
 domains of observations.
 }\label{fig:atv}
\end{figure*}

Figure~\ref{fig:atv} shows the surface abundances of C, O, Mg,
and Fe with respect to effective temperature in models computed with
$D_{\rm R}=0.0,\, 0.5,\, 1.0,\, 5.0,\, 9.0$, together with the observations of
Varenne \& Monier~(\cite{vm99}).
For main sequence stars, with effective temperatures between
$6\,600$\,K$\leq T_{\rm eff}\leq 7\,800$\,K,
the depletions of abundances are decreasing with respect to the radiative
diffusivity parameter.
With $D_{\rm R}\cong1$ the depletions amount to about a factor
$\approx 2-3$ for all chemical species;
that corresponds to the observations even if, for oxygen, the distribution
 of the observed abundances are rather scattered,
 probably resulting from non-{\sc lte} effects or 3D temperature inhomogeneities
 in the atmosphere (Asplund et al.~\cite{acgk00}).

Taking into account observational error bars, Fig.~\ref{fig:atv} shows that
the best agreement is obtained for $D_{\rm R}\cong1$.

\section{Comments on stellar models with $D_{\rm R}=1$}
\label{sec:ill}
Using stellar models computed with element diffusion in the following
 we give two illustrations of the use of radiative diffusivity
(1) to estimate the age of Hyades
and (2) to give properties of a sample of Pop~I stellar models.

\begin{figure}
\psfig{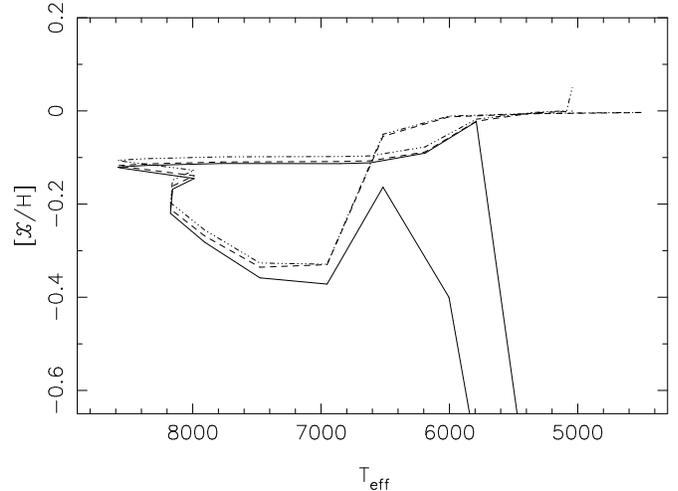}
\caption[]{Profiles of expected surface abundances
(in dex) with respect to effective temperature for ${\cal X}=\rm He$ (dashed),
Li (full) and N (dash-dot-dot-dot-dash) for a sample of
 ``Hyades'' stellar models at the age of $570$\,My computed with
$D_{\rm R}=1.0$.
}\label{fig:pre}
\end{figure}

\begin{figure}
 \centerline{
 \psfig{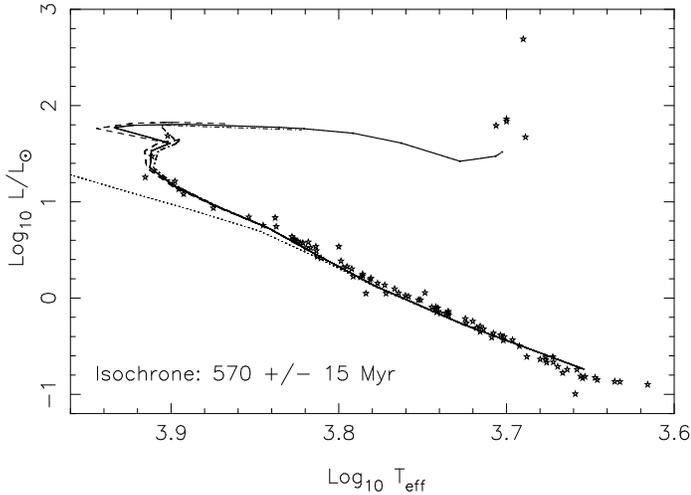}}
 \caption[]{Isochrones of
 ZAMS (dotted), 570\,Myr (full),  585\,Myr (dash-dot-dash)
 and 555\,Myr (dashed) for the Hyades open cluster
 computed with element diffusion and $D_{\rm R}=1$.
 }\label{fig:isob}
\end{figure}

\paragraph{Age and predicted abundances of the Hyades stars:}
So far we have used for the Hyades open cluster the age and the initial
chemical composition derived from the work of
Lebreton et al.~(\cite{lfl01}) that does not take into account
 the element diffusion. Using stellar
models computed with $D_{\rm R}=1.0$, and the corrected initial helium
content and metallicity introduced in Sect.~\ref{sec:hay},
we have reestimated the age by fitting isochrones to the observations
of Balachandran~(\cite{b95}).
Figure~\ref{fig:isob} shows the isochrones.

The inclusion of the element diffusion leads to the age
$t_{\rm Hyades}=570\pm15$\,Myr, i.e. an  increase of $\approx 3\%$
with respect to the result of Lebreton et al.~(\cite{lfl01}).
As already emphasized by Lebreton et al.~(\cite{lfl01}) one can see
 the sensitivity of the age to the loci of the two components of the
$\theta$\,Tau binary system with $\theta^2$\,Tau located right on the turn-off.

Figure~\ref{fig:pre} reproduces the predictions of surface
abundances for helium, lithium and nitrogen with 
respect to the effective temperature. Below $T_{\rm eff}\loa 6\,000$\,K the
external convective zone extends deeply enough to reach the area where
lithium is nuclearly destroyed, resulting in a low surface abundance.
For $6\,400\loa T_{\rm eff}\loa 6\,600$\,K the lithium abundance is close to
its initial value as for the other elements. 
Above $\goa 6\,600$\,K and until $\approx 8\,200$\,K 
the depletion due to the element diffusion is more and more counter
balanced by the mixing generated by the increasing radiative diffusivity.
For stars above the turn-off, first the evolutionary time scale is small and 
the element diffusion is less efficient and second,
the mixing resulting from the dredge up becomes more and more efficient and
the surface abundances keep about their initial values.

Figure~\ref{fig:Li} shows the 
surface abundance of lithium with respect to effective temperature.
The so-called Li dip (Boesgaard \& Tripicco~\cite{bt86}), located 
in the vicinity of $T_{\rm eff}~\approx 6\,500$\,K is not reproduced by our
models. As in the standard solar model, see Table~\ref{tab:sol}, 
the radiative diffusivity combined with element diffusion
does not suffice to explain the lithium depletion. As clearly demonstrated by
Charbonnel et al.~(\cite{cvm94}) the Li dip does not result from the
gravitational settling; according
to the present paradigm (Sills \& Deliyannis~\cite{sd00}), it results
from the mixing
generated beneath the tachocline due to the shear caused by the
differences of rotational velocities between the convection zone and the
radiative interior (Talon \& Charbonnel~\cite{tc98}).
At the age under consideration, according to
the rotational status of the star and to the extend of the external convection
zone, the mixed region extends towards the part of the envelope
where the lithium is nuclearly destroyed.
The other elements, except beryllium and boron, being nuclearly
destroyed in the core are not affected by this process.
The radiative diffusivity then acts together with the other processes to
inhibit the gravitational settling, generating the mild depletion of
metals as observed by Varenne \& Monier~(\cite{vm99}).

\medskip
For reference, Table~\ref{tab:ex} shows 
theoretical surface parameters of our ``Hyades''  
stellar models and reveals significant changes of
 the surface abundances according to the stellar mass. Therefore,
at present, neither the helium mass fraction nor the metallicity
have a unique value at the surface of the main sequence star members
of the cluster.

\begin{table*}
\caption[]{Typical physical parameters of ``Hyades''  
stellar models computed with $D_{\rm R}=1$ at age $570$\,Myr.
$M_\star$, $T_{\rm eff}$, $\log g$ and $M_{\rm bol}$ are respectively
the mass,
the effective temperature (in K), the logarithm of the gravity and the
bolometric magnitude ($M_{\rm bol\odot}=4.75$). $\rm [\frac{He}{H}]$, 
$\rm [\frac CH]$, $\rm [\frac NH]$, $\rm [\frac OH]$,
$\rm [\frac{Mg}H]$, 
$\rm [\frac{Fe}H]$ are the relative surface abundances with
respect to the Hyades one: $+0.14$\,dex.
}\label{tab:ex}
\centerline{
\begin{tabular}{llllllllllllll} \hline \\
$M_\star/M_\odot$  & $T_{\rm eff}$ & $\log g$ & $M_{\rm bol}$ & $\rm [\frac{He}H]$ & 
$\rm [\frac CH]$ & $\rm [\frac NH]$ & $\rm [\frac OH]$ & $\rm [\frac{Mg}H]$ &
 $\rm [\frac{Fe}H]$\\
\\  \hline \\
 $0.8$ & $4\,508$ & $4.65$ & $6.59$  & $-0.003$ & $-0.002$ & $-0.002$ & $-0.002$ & $-0.002$ & $-0.003$ \\
 $1.0$ & $5\,324$ & $4.57$ & $5.43$  & $-0.004$ & $-0.004$ & $-0.004$ & $-0.004$ & $-0.004$ & $-0.004$ \\
 $1.2$ & $6\,003$ & $4.46$ & $4.44$  & $-0.009$ & $-0.008$ & $-0.008$ & $-0.008$ & $-0.008$ & $-0.009$ \\
 $1.4$ & $6\,517$ & $4.34$ & $3.61$  & $-0.040$ & $-0.035$ & $-0.037$ & $-0.040$ & $-0.033$ & $-0.038$ \\
 $1.6$ & $6\,953$ & $4.24$ & $2.93$  & $-0.259$ & $-0.279$ & $-0.274$ & $-0.252$ & $-0.226$ & $-0.293$ \\
 $1.8$ & $7\,477$ & $4.18$ & $2.35$  & $-0.264$ & $-0.276$ & $-0.270$ & $-0.255$ & $-0.252$ & $-0.317$ \\
 $2.0$ & $7\,908$ & $4.12$ & $1.84$  & $-0.208$ & $-0.208$ & $-0.207$ & $-0.199$ & $-0.190$ & $-0.234$ \\
 $2.2$ & $8\,175$ & $4.03$ & $1.37$  & $-0.161$ & $-0.156$ & $-0.156$ & $-0.152$ & $-0.144$ & $-0.173$ \\
 $2.4$ & $8\,159$ & $3.89$ & $0.927$ & $-0.122$ & $-0.156$ & $-0.115$ & $-0.134$ & $-0.108$ & $-0.127$ \\
 $2.5$ & $7\,991$ & $3.79$ & $0.721$ & $-0.104$ & $-0.098$ & $-0.098$ & $-0.097$ & $-0.093$ & $-0.109$ \\
$2.55$ & $8\,592$ & $3.76$ & $0.323$ & $-0.088$ & $-0.080$ & $-0.080$ & $-0.079$ & $-0.070$ & $-0.083$ \\
$2.57$ & $7\,823$ & $3.57$ & $0.236$ & $-0.082$ & $-0.074$ & $-0.074$ & $-0.074$ & $-0.065$ & $-0.077$ \\
$2.59$ & $5\,151$ & $3.22$ & $1.169$ & $-0.002$ & $-0.006$ & $-0.004$ & $-0.001$ & $-0.001$ & $-0.001$ \\
\\  \hline \\
\end{tabular}
}
\end{table*}
 \begin{figure*}
 \psfig{figure=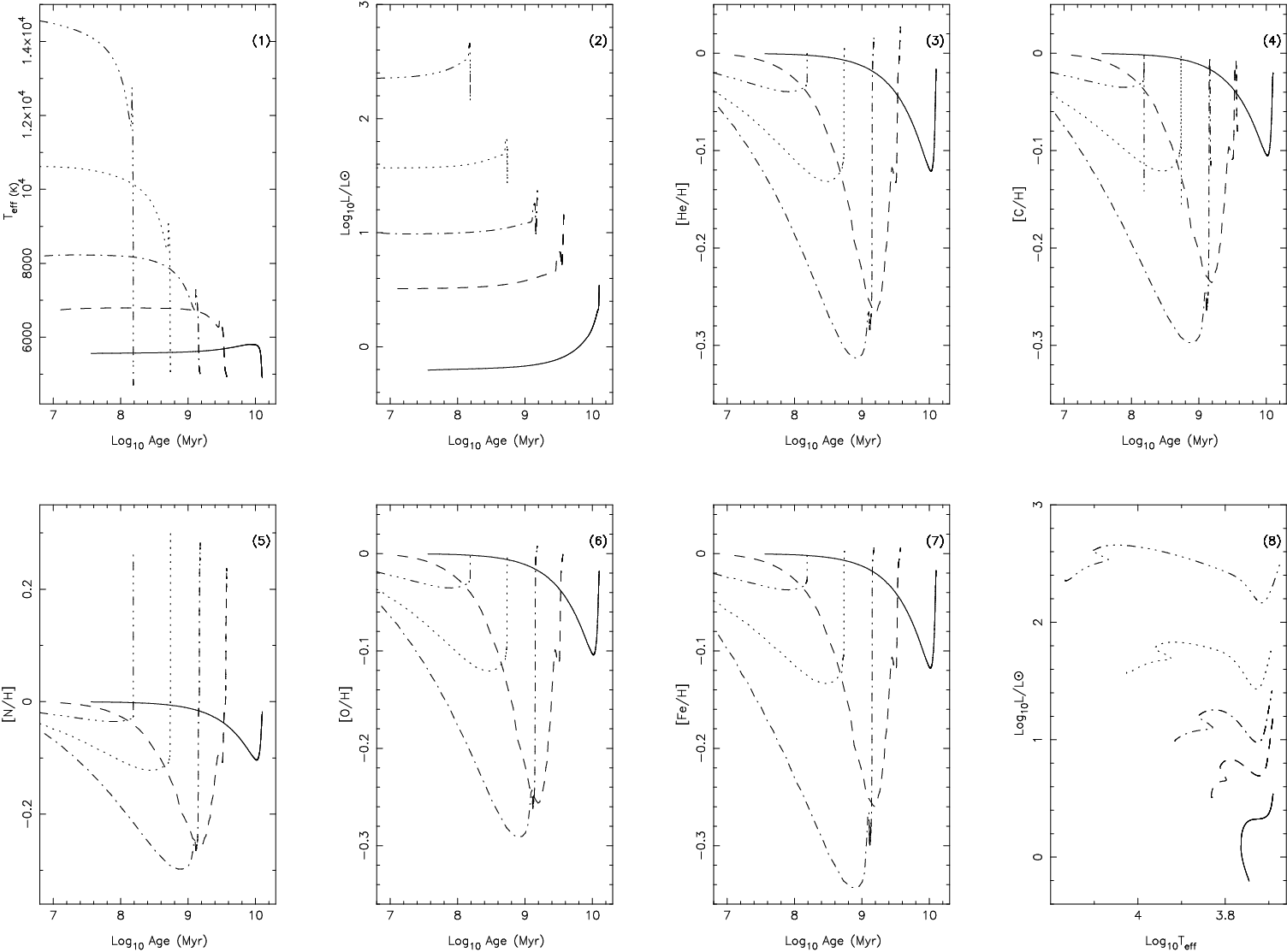,width=18.cm,angle=270}
 \caption[]{
 Panels (1) to (7) show the
 profiles, with respect to age, of the effective temperature,
 of the luminosity, of abundance mass
 fractions at the surface of a sample of Pop. I stellar models
  computed with $D_{\rm R}=1.0$.
Panel (8) shows the corresponding HR diagram. The  labels are self explanatory.
The stellar masses are respectively
 $M_\star/M_\odot=1.0$ (full), $1.4$ (dashed),
 $1.8$ (dot-dash-dot-dash), $2.5$ (dotted),
$4.0$ (dash-dot-dot-dot-dash).
 }\label{fig:F}
\end{figure*}

\paragraph{A sample of Pop. I stellar models:} 
To investigate %glance attention at 
the change of the surface
abundances resulting from the element diffusion on the main sequence
and subgiant stars we have computed typical evolutions
of Pop. I stellar models for respectively
$1.\,M_\odot$, $1.4\,M_\odot$, $1.8\,M_\odot$, $2.5\,M_\odot$ and $4.\,M_\odot$.
Models are initialized at homogeneous {\sc zams} with solar composition.
Figure~\ref{fig:F} shows, with respect to age, the
profiles of the surface abundance and the corresponding HR diagrams.
Along the main sequence,
the surface mass fractions of helium and metals
are depleted, but along the giant branch, the surface
abundances recover about their initial values as a consequence
of the dredge-up. Therefore it not possible to derive a unique
metallicity for all stats of an open cluster like Hyades.

Note that for the model of $1\,M_\odot$ at $4.65$\,Gyr, the age of the Sun,
we obtain a depletion for iron of $\approx 0.059$\,dex, consistent with the present day
adopted difference between photospheric and meteoritic observed abundances
(Asplund et al.~\cite{ants00}).
  
\begin{figure}
\centerline{
\psfig{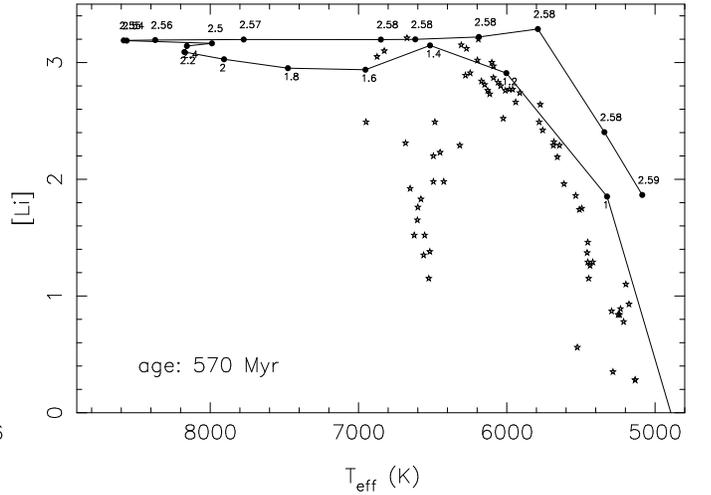}
}
\caption{
Surface abundance of lithium with respect to effective temperature of stellar
models (full)
of the Hyades computed with $D_{\rm R}=1.0$ at the age of
$570$\,Myr. The labels of dots give the mass (solar unit) of models.
Observed lithium abundances and effective temperatures
of the Hyades field stars ($\star$) are taken from Balachandran~(\cite{b95}). 
}\label{fig:Li}
\end{figure}

\section{Discussion}\label{sec:con}
Radiative diffusivity is a basic physical phenomenon that complements standard
element diffusion in stellar modeling.
In this exploratory work, we have introduced in the calculation of the
microscopic diffusion coefficients, an estimate of
the radiative diffusivity resulting from photon-ion collisions
that acts predominantly in the external layers of stars
of high effective temperature.
This simple treatment needs an efficiency factor, the radiative
diffusivity parameter $D_{\rm R}$. We have constrained 
$D_{\rm R}$ by the fit of observations chosen independent of rotation. 
The studies presented in Sect.~\ref{sec:estim} strongly suggest
that a value close to unity, $D_{\rm R}=1.0^{+2.0}_{-0.2}$,
satisfies all the observational constraints.
We have adopted a unique value $D_{\rm R}=1$
for all chemical element species and in all physical conditions.
 
Without taking into account other physical processes that inhibit the
gravitational sedimentation, namely
mass-loss, or turbulent mixing generated by rotation or
by some other turbulent process,
we find that the implementation of the radiative diffusivity inhibits
the large gravitational
settling resulting from the element diffusion in outer layers of
early-type star models with masses
larger than $\ga 1.4M_\odot$.

The radiative diffusivity is not able to solve all the open
questions of element depletion in stellar models
(e.g. Balachandran~\cite{b95}; Sills \& Deliyannis~\cite{sd00} for reviews).
Here, our goal is to emphasize
on its large magnitude and to point out that it should be added to the other
mixing processes. Radiative diffusivity simply acts together with the mixing
induced by rotation (Zahn~\cite{z92}), the
mass-loss (Chaboyer et al.~\cite{cdg99}), the turbulent mixing
(Schatzman~\cite{s69}) and the radiative
accelerations (Alecian et al.~\cite{amt89}) 
to inhibit the large element diffusion.

The implementation of the radiative diffusivity in stellar modeling is
necessary, in particular, for the calculation of:
\begin{itemize}
\item precise models of intermediate mass stars that are
targets of asteroseismological experiments like {\sc corot} or {\sc eddington},
\item theoretical isochrones
including element diffusion, for young cluster
presenting stars with masses above $1.4\,M_\odot$ at the turn-off.
\end{itemize}

Our simplified model takes into account only the mixing
generated by the anisotropic
component of the radiative diffusion; 
it ignores the radiative acceleration
and so cannot describe the AmFm phenomenon
(Michaud et al.~\cite{mtcp83}; Alecian~\cite{a96};
 Hui-Bon-Hua \& Alecian~(\cite{ha98}; Richer et al.~\cite{rmt00}).

We are aware that our estimation of the radiative diffusivity
is only a first approximation; owing to its efficiency
we are confident of the fact that the diffusivity generated by
the radiative collisions is a part of
the missing physical process responsible for the too large efficiency of the
microscopic diffusion coefficient.
Our description of the radiative diffusivity is
phenomenological and preliminary and needs further extended developments based
on the kinetic theory of gases and radiation.

\begin{acknowledgements}
We are grateful to A. Baglin for many valuable suggestions and advice
concerning the concept of this paper.
Fruitful discussions with J.-P. Zahn, Y. Lebreton, G. Alecian,
M.-J. Goupil and R. Monier are gratefully acknowledged.
We would like to express our thanks to J. Provost for the
computation of the small differences of the calibrated solar models.
The comments of the anonymous referee significantly improved the presentation
of this paper.
This research has made use of the Simbad data base, operated at
CDS, Strasbourg, France.
This work has been performed using the computing facilities 
provided by the OCA program
``Simulations Interactives et Visualisation en Astronomie et M\'ecanique 
(SIVAM)''.
\end{acknowledgements}


\begin{thebibliography}{}

\bibitem[1989]{amt89}
Alecian G., Michaud G., Tully J., 1989, A\&A 411, 882

\bibitem[1996]{a96}
Alecian G., 1996, A\&A 310, 872

\bibitem[1994]{af94}
Alexander D.R., Fergusson J.W., 1994, ApJ 437, 879
  
\bibitem[1989]{ag89}
Anders E., Grevesse N.,  1989, Geochimica et Cosmochimica Acta 53, 197

\bibitem[1999]{aar99}
Angulo C., Arnould M., Rayet M.,
and the NACRE collaboration, 1999, Nuclear Physics A 656, 1, 
and WEb site http://pntpm.ulb.ac.be/Nacre/nacre.htm

\bibitem[2000a]{acgk00}
Asplund M., Carlsson M., Garcia-Peres A.E., Kiselman D., 2000,
Oxygen Abundances in Old Stars and Implications to Nucleosynthesis and
Cosmology. In: 24th meeting of the IAU, Joint
Discussion 8, August 2000, Manchester, England.

\bibitem[2000b]{ants00}
Asplund M., Nordlund \AA., Trampedach R., Stein R.F., 2000, A\&A 359, 743

\bibitem[1972]{b72}
Baglin A., 1972, A\&A 19, 45 

\bibitem[1995]{b95}
Balachandran S., 1995, ApJ 446, 203

\bibitem[1995]{ba95}
Basu S., Antia H.M., 1995, MNRAS 276, 1402

\bibitem[1997]{b97}
Basu S., 1997, The Seismic Sun. In: Provost J., Schmider F.X, (eds)
Sounding solar and stellar interiors. IAU Symposium 181,
Kluwer Academic Publisher, p. 137

\bibitem[1995]{bbg95}
Bodmer R., Bochsler P., Geiss J., et al. 1995, Space Sci. Rev. 72, 61

\bibitem[1986]{bt86}
Boesgaard A.M., Tripicco M., 1986, ApJ 302, 249

\bibitem[1958]{b58}  
B\"{o}hm-Vitense E., 1958, Z. Astrophys. 54, 114

\bibitem[1998]{btz99}
Brun S., Turck-Chi\`eze S., Zahn J.P., 1999, ApJ 525, 1032

\bibitem[1969]{b69}
Burgers J.M., 1969, Flow equations for composite gases,
Academic Press, New york and London

\bibitem[1999]{clm99}
Cayrel R., Lebreton Y., Morel P., 1999.
Survival of $\rm ^6Li$ and $\rm ^7Li$ in metal-poor stars. In: M. Spite (ed.)
Galaxy Evolution: Connecting the Distant Universe
with the Local Fossil Record, Astrophys. Space Sci. Lib.,
v. 265, Issue 1/4, p. 87

\bibitem[1999]{cdg99}
Chaboyer B., Demarque P., Guenther D.B., 1999, ApJ 525, L41

\bibitem[2001]{cfn01}
Chaboyer B., Fenton W.H., Nelan J.E., et al., 2001, ApJ 562, 521

\bibitem[1994]{cvm94}
Charbonnel C., Vauclair S., Maeder A., et al., 1994, A\&A 283, 155

\bibitem[1996]{cda96}
Christensen-Dalsgaard J., D\"appen W., Ajukov  S.V.,
et al., 1996, Science 272,1286

\bibitem[1989]{cgk89}
Cox A.N., Guzik J.A., Kidman R.B., 1989, ApJ 342, 1187

\bibitem[1994]{cl94}
Cunha K., Lambert D.L., 1994, ApJ 426, 170 

\bibitem[2000]{dcbs00}
Daflon S., Cunha K., Becker S.R., Smith V.V., 2001, ApJ 552, 309

\bibitem[1973]{eff73}
Eggleton P.P., Faulkner J., Flannery B.P., 1973, A\&A 23, 325

\bibitem[1997]{g97}
Gabriel M., 1997, A\&A 327, 771

\bibitem[1997]{gm97}
Gautier D., Morel P., 1997, A\&A 323, L9

\bibitem[1972]{gr72}
Geiss J., Reeves H., 1972, A\&A 18, 126

\bibitem[1993]{gn93}
Grevesse N., Noels A., 1993, Cosmic Abundances of the Elements.
In: Prantzos N., Vangioni-Flam E. \& Cass\'e M. (eds.)
Origin and Evolution of the Elements. Cambridge University Press, 14

\bibitem[1997]{gtl97}
Grec G., Turck-Chi\`eze S., Lazrek M., et al., 1997, {\sc golf} results;
to-day's view on the solar modes. In: Provost J., Schmider F.X. (eds) Sounding
solar and stellar interiors. {\sc iau} syposium 181, Kluwer Academic Publisher,
91

\bibitem[2000]{hlw00}
Heger A., Langer N., Woosley S.E., 2000, ApJ 528, 368

\bibitem[1998]{ha98}
Hui-Bon-Hoa A., Alecian G., 1998, A\&A 332, 224 

\bibitem[1999]{lpcbf99}  
Lebreton Y., Perrin M.-N., Cayrel R., Baglin A.,
Fernandes J., 1999, A\&A 350, 587

\bibitem[2001]{lfl01}
Lebreton Y., Fernandes J., Lejeune T., 2001, A\&A 374, 540

\bibitem[1983]{mtcp83}
Michaud G., Tarasick D., Charland Y., Pelletier C., 1983, ApJ 269, 239

\bibitem[1993]{mp93}
Michaud G., Proffitt C.R., 1993, Particule Transport Process.
In: Baglin A., Weiss W.W. (eds.) Inside the Stars, IAU Colloquium 137,
ASP Conference Series, Vol. 40, 246

\bibitem[1978]{m78}
Mihalas D., 1978, Stellar Atmospheres, 2d Ed. Freeman and Cie.

\bibitem[1984]{mm84}
Mihalas D., Weibel-Mihalas B., 1984, Foundations of Radiation Hydrodynamics,
Oxford University Press, p. 461-472

\bibitem[1930]{m30}
Milne E.A., 1930, Quart. J. of Math. (Oxford), 1, 1

\bibitem[1997]{m97}
Morel P., 1997, A\&AS 124, 597

\bibitem[1997]{mpb97}
Morel P., Provost J., Berthomieu G., 1997, A\&A 327, 349

\bibitem[1999]{mb99}
Morel P., Baglin A., 1999, A\&A 345, 156

\bibitem[2000]{mpltb00}
Morel P., Provost J., Lebreton Y., Th\'evenin F., Berthomieu G., 2000,
A\&A 363, 675

\bibitem[1986]{ppfm86}
Paquette C., Pelletier C., Fontaine G., Michaud G., 1986, ApJS 61, 177

\bibitem[1991]{pm91}
Proffit C.R., Michaud G., 1991, ApJ 371, 584

\bibitem[1996]{rvcd96}
Richard O., Vauclair S., Charbonnel C., Dziembowsky D.A., 1996, A\&A 312, 1000

\bibitem[2000]{rmt00}
Richer J., Michaud G., Turcotte S., 2000, ApJ 529, 338

\bibitem[1996]{rsi96}  
Rogers F.J., Swenson F.J., Iglesias C.A., 1996, ApJ 456, 902

\bibitem[2001]{sw01}
Salaris M., Weiss A., 2001, A\&A 376, 955

\bibitem[1969]{s69}
Schatzman E., 1969, A\&A 3, 331

\bibitem[1977]{s77}
Schatzman E., 1977, A\&A 56, 211

\bibitem[2000]{szm00}
Schatzman E., Zahn J.-P., Morel P., 2000, A\&A 364, 876

\bibitem[2000]{sd00}
Sills A., Deliyannis C.P., 2000, ApJ 544, 944

\bibitem[1998]{tc98}
Talon S., Charbonnel C., 1998, A\&A 335, 959

\bibitem[1930]{t30}
Thomas L.H., 1930, Quart. J. of Math. (Oxford), 1, 239

\bibitem[1994]{tbl94}
Thoul A.A., Bahcall J.N., Loeb A., 1994, ApJ 421, 828

\bibitem[1997]{tbb97}
Turck-Chi\`eze S., Basu S., Brun S., et al., 1997, Solar Phys. 175, 247

\bibitem[1998a]{trmir98}
Turcotte S., Richer J., Michaud G., Iglesias C.A., Rogers F.J., 1998,
ApJ 504, 539

\bibitem[1998b]{trm98}
Turcotte S., Richer J., Michaud G., 1998b, ApJ 504, 559

\bibitem[1999]{vm99}
Varenne O., Monier R., 1999, A\&A 351, 247

\bibitem[1992]{z92}
Zahn J.-P., 1992, A\&A 265, 115

\end{thebibliography}
\end{document}